\documentclass[11pt]{article}
\usepackage[margin=1in]{geometry} 

\usepackage{amsmath,amssymb,algorithm,algorithmic,theorem,float,bbm,bm,enumerate}

\usepackage[font=small,labelfont=sc]{caption}
\usepackage{subcaption}
\usepackage{url}
\usepackage{graphicx}
\usepackage{enumitem}
\usepackage[colorlinks, citecolor=black, urlcolor=black, linkcolor=black]{hyperref}
\usepackage{setspace}
\usepackage[round]{natbib}

\DeclareMathOperator{\argmin}{argmin}

\begin{document}

\title{Disciplining Deliberation: A Sociotechnical Perspective on Machine Learning Trade-offs\thanks{Accepted for publication in the \textit{British Journal for the Philosophy of Science}.}}

\author{
Sina Fazelpour \\ 
Northeastern University \\ 
s.fazel-pour@northeastern.edu
}

\date{}

\sloppy

\maketitle

\begin{abstract}
This paper examines two prominent formal trade-offs in artificial intelligence (AI)---between predictive accuracy and fairness, and between predictive accuracy and interpretability. These trade-offs have become a central focus in normative and regulatory discussions as policymakers seek to understand the value tensions that can arise in the social adoption of AI tools. The prevailing interpretation views these formal trade-offs as directly corresponding to tensions between underlying social values, implying unavoidable conflicts between those social objectives. In this paper, I challenge that prevalent interpretation by introducing a sociotechnical approach to examining the value implications of trade-offs. After providing a set-theoretic understanding of these model trade-offs, I argue that their implications cannot be understood in isolation from the technical, psychological, organizational, and social factors that shape the integration of AI models into decision-making pipelines. Specifically, I identify three key considerations---validity and instrumental relevance, compositionality, and dynamics---for contextualizing and characterizing these implications. These considerations reveal that the relationship between model trade-offs and corresponding values depends on critical choices and assumptions. Crucially, judicious sacrifices in one model property for another can, in fact, promote both sets of corresponding values. The proposed sociotechnical perspective thus shows that we can and should aspire to higher epistemic and ethical possibilities than the prevalent interpretation suggests, while offering practical guidance for achieving those outcomes. Finally, I draw out the broader implications of this perspective for AI design and governance, highlighting the need to broaden normative engagement across the AI lifecycle, develop legal and auditing tools sensitive to sociotechnical considerations, and rethink the vital role and appropriate structure of interdisciplinary collaboration in fostering a responsible AI workforce.
\end{abstract}

\section{Introduction}\label{sec:intro}
Our aims and values are diverse and many. So, unsurprisingly, in many cases we can't have them all, as interventions that realize some will sacrifice others. In contending with this reality, we thus regularly face difficult questions about value trade-offs. To navigate these trade-offs, formal analyses of decision scenarios offer crucial guidance, helping decision-makers identify where such conflicts arise and consider how best to resolve them. When incorporated appropriately into practical reasoning, these formal frameworks bring a valuable level of rigor and discipline to deliberations.

This paper examines two highly publicized formal trade-offs in the field of responsible artificial intelligence (AI)---the trade-off between \emph{predictive accuracy} and \emph{fairness}, and between \emph{predictive accuracy} and \emph{interpretability}. 
Ideally, the social and institutional adoption of AI tools should enhance decision-making by promoting key societal values, including accuracy, validity and reliability, fairness, transparency, trust, and safety. However, as these trade-offs suggest, it may be inherently impossible to simultaneously optimize AI models for promoting formal specifications of all these values. Ensuring fairness (in some statistical sense) might require sacrificing some level of predictive accuracy~\citep{kleinberg2018inherent, corbett2017algorithmic}. Similarly, the most accurate models might be ``blackboxes'' that lack interpretability (in some sense), threatening values such as trust, safety, or procedural fairness, which interpretability is said to support~\citep{murdoch2019definitions,molnar2022}. 

As policymakers seek to better understand value tensions that can emerge in responsible AI governance,\footnote{For example, the U.S. National Institute of Standards and Technology’s influential AI risk management framework calls on researchers to provide ``Guidance on the tradeoffs and relationships that may exist among trustworthiness characteristics'', where trustworthiness characteristics includes values such as validity and reliability, fairness, and interpretability.} 
these formal trade-offs have become a core locus of normative and regulatory engagement~\citep{fleisher2022understanding,london2019artificial,kearns2019ethical,babic2021beware,rudin2019stop,tabassi2023artificial,johnson2021algorithmic,loi2021choosing,fazelpour2021algorithmicbias,sullivan2022understanding}. 
Specifically, questions arise as to how to understand these trade-offs and how to interpret their implications for the relationships---and potential conflicts---between corresponding social values. 

As I explain in Section~\ref{sec:trade-offs}, at a high level of abstraction, both trade-offs can be viewed as the result of learning AI models under non-trivial constraints that encode objectives beyond maximizing predictive accuracy. These formal trade-offs emerge when interventions designed to enforce such constraints prevent access to the most accurate models. A common interpretation holds that there is a \textit{direct correspondence} between these interventions that lead to formal trade-offs and their impacts on underlying values. That is, according to view, the formal trade-offs express unavoidable tensions between competing values. From this perspective, making reasoned decisions requires that interventions in the model space directly reflect our value priorities. In Section~\ref{sec:trade-offs}, I describe this common interpretation and explain how it shapes the disciplinary organization and epistemic culture of responsible AI communities.

In this paper, I offer an alternative, \textit{sociotechnical} approach to understanding the value implications of formal trade-offs, one that challenges their common interpretation. As emphasized by an emerging body of work, understanding the social impacts of AI models from this perspective requires evaluating not only the properties of AI models in isolation, but also the technological, psychological, and social processes that shape their design, development, and deployment~\citep{selbst2019fairness,suresh2019framework,fazelpour2020algorithmic}. In Section~\ref{sec:against}, I show how adopting such a sociotechnical perspective complicates the translation of formal, model-level trade-offs into interrelations between values. Specifically, I introduce three key considerations for contextualizing and interpreting the implications of these trade-offs: \textit{validity and instrumental relevance} (Section~\ref{sec:validity}), \textit{compositionality} (Section~\ref{sec:compositionality}), and \textit{dynamics} (Section~\ref{sec:dynamics}). These considerations demonstrate that the relationship between formal model-level trade-offs and corresponding values is not straightforward and rests on many critical assumptions. Importantly, a closer examination of these assumptions shows that interventions sacrificing one formal property for another at the model level need \textit{not} entail a cost for the corresponding value, and can, in fact, promote \textit{both} values in practice.

This sociotechnical approach to interpreting trade-offs carries significant implications for the scope and mode of normative engagement in AI evaluation, legal and policy debates in AI governance, and the organizational structuring of responsible AI workforce. I draw out these implications in Section~\ref{sec:discussion}. To be clear, adopting this lens does not eliminate the need to navigate value tensions. Indeed, as I discuss, it often brings into sharp focus new value trade-offs. Nor does it suggest that existing normative discussions on formal trade-offs are not valuable. What the sociotechnical shows is that, in many cases, we \textit{can} and \textit{should} have higher aspirations than the common interpretation would suggest. Realizing these better epistemic and ethical possibilities requires adopting a sociotechnical lens that broadens the scope and deepens the level of normative engagement with AI technologies, rethinking and expanding the tools of auditing, and appreciating the vital role of interdisciplinary collaboration in fostering responsible AI. 

\section{Trade-offs in AI-based Decision-making}\label{sec:trade-offs}
This section introduces the intuition behind the two formal trade-offs and explains the common interpretation of their value implications. It is worth noting that discussions of both trade-offs predate recent AI applications. The accuracy-fairness trade-off, for example, has been discussed in economics~\citep{young1994equity} and education~\citep{willingham2013gender}. And the accuracy-interpretability trade-off has been discussed by statisticians~\citep{plate1999accuracy}. 

\subsection{Prediction-based decision-making and the cost of constraints}\label{sec:prediction-based}
To ground the discussion, consider a typical binary classification task where the goal is to learn the relationship between a set of input features $X$ and a target label $Y$, with $X$ drawn from $\mathcal{X}$ and $Y$ from $\mathcal{Y} = \{0, 1\}$. Once this relationship is learned, given a feature vector $x \in \mathcal{X}$, we can then infer the corresponding label $y \in \{0, 1\}$, and use this prediction to inform or drive a relevant decision.  For example, in content moderation, $X$ could represent information about social media posts (including metadata), and $Y$ could denote whether a post contains offensive language. Supervised learning algorithms facilitate this process by mapping datasets of past observations to predictive models (or hypotheses) that can be used to make inferences about new cases. That is, given a dataset of labeled examples, these algorithms output a predictive model $h: \mathcal{X} \rightarrow \{0,1\}$, from a set of possible predictive models $\mathcal{H}$. Typically, these algorithms are designed to find a model $h^* \in \mathcal{H}$ that optimizes predictive performance according to some evaluation function, such as minimizing a loss function like empirical risk $\mathcal{\hat{L}}$:
$$h^* = \argmin_{h\in\mathcal{H}}\mathcal{\hat{L}}(h)$$

In practice, there is often a multiplicity of models that perform comparably well with respect to this optimization objective~\citep{coston2021characterizing,marx2020predictive,d2021revisiting}.\footnote{In the literature similar phenomenon---to the one referred to here as \emph{the set of accurate models}---are called the Rashomon set~\citep{breiman2001statistical}, model multiplicity~\citep{marx2020predictive},  underspecification~\citep{d2020underspecification}, and the set of good models~\citep{coston2021characterizing}.} 
Let us define \emph{the set of accurate models}, $\mathcal{H}_A$, as the set of models that exhibit, at their worst, only $\epsilon$ more loss than the best predictive model, $\hat{\mathcal{L}}(h^*)$: 

$$\mathcal{H}_A := \{h\in \mathcal{\mathcal{H}}: \hat{\mathcal{L}}(h) \le \hat{\mathcal{L}}(h^*) + \epsilon\}$$

Using this set theoretic framing, we can understand trade-offs by viewing the act of \emph{enforcing} fairness or interpretability as an intervention that imposes a non-trivial constraint on the set of available predictive models. In doing so, it \textit{restricts} access to a subset of models in $\mathcal{H}$ that may not intersect with the set of accurate models, $\mathcal{H}_A$.

In particular, let $\mathcal{H}_C \subseteq \mathcal{H}$ refer to a \emph{subset} of possible predictive models that satisfy some constraint(s). Constraining model choice in this way may come with a predictive cost, as the best models within the restricted set $\mathcal{H}_C$ can, at best, only match the predictive performance of models from the broader hypothesis class $\mathcal{H}$. Formally:

$$\text{Since } \mathcal{H}_C \subseteq \mathcal{H} \text{, then } \min_{h\in \mathcal{H}_C}\mathcal{\hat{L}}(h) \ge \min_{h\in \mathcal{H}}\mathcal{\hat{L}}(h) $$

\begin{figure}
     \centering
     \begin{subfigure}[b]{0.32\textwidth}
         \centering
         \includegraphics[width=\textwidth]{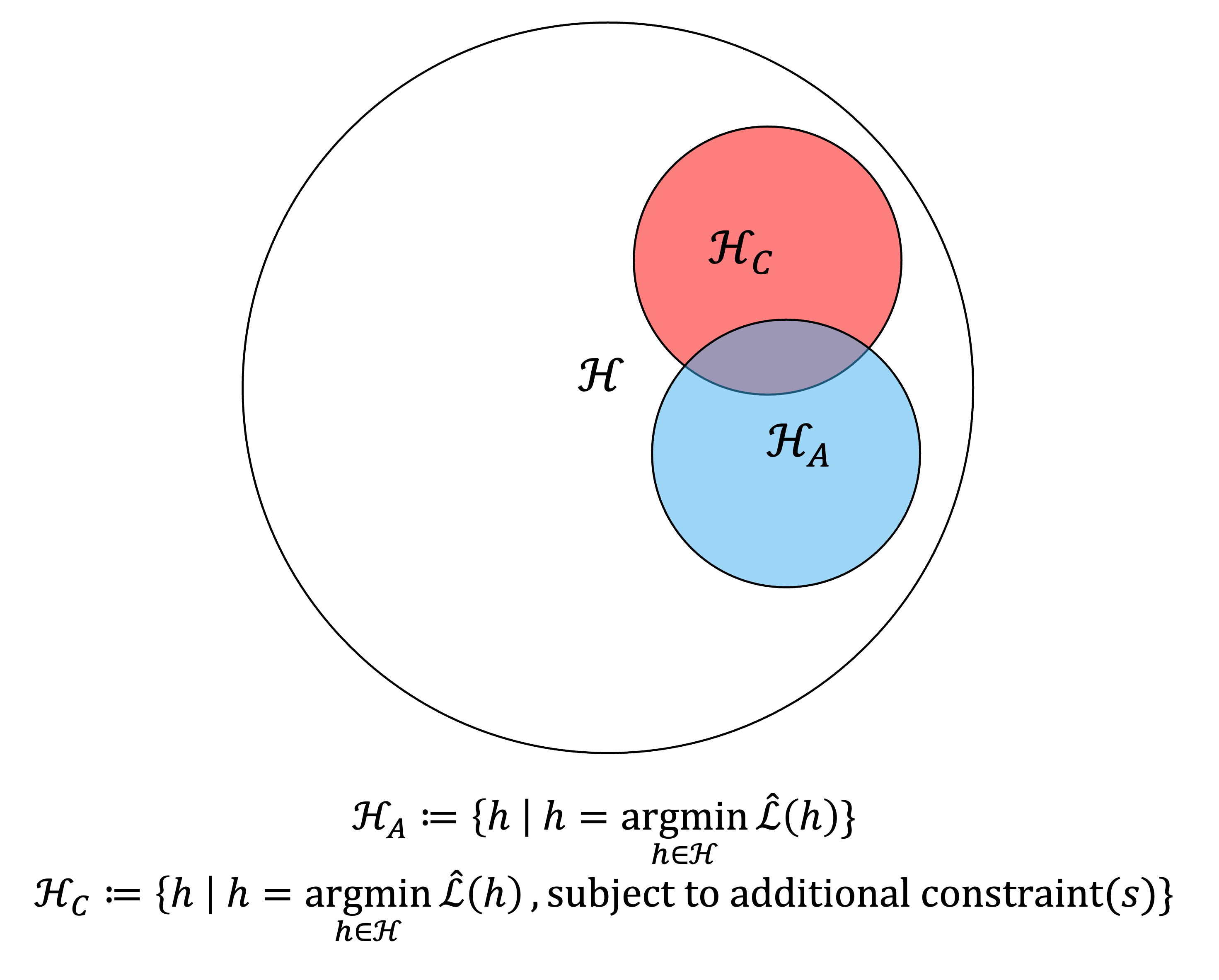}
         \caption{}
         \label{fig:accuracy-constraint}
     \end{subfigure}
     \hfill
     \begin{subfigure}[b]{0.32\textwidth}
         \centering
         \includegraphics[width=\textwidth]{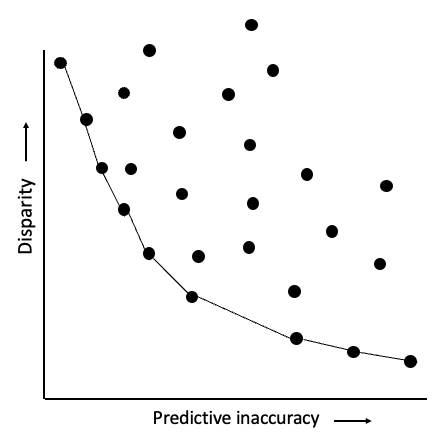}
         \caption{}
         \label{fig:accuracy-fairness-pareto}
     \end{subfigure}
     \hfill
     \begin{subfigure}[b]{0.32\textwidth}
         \centering
         \includegraphics[width=\textwidth]{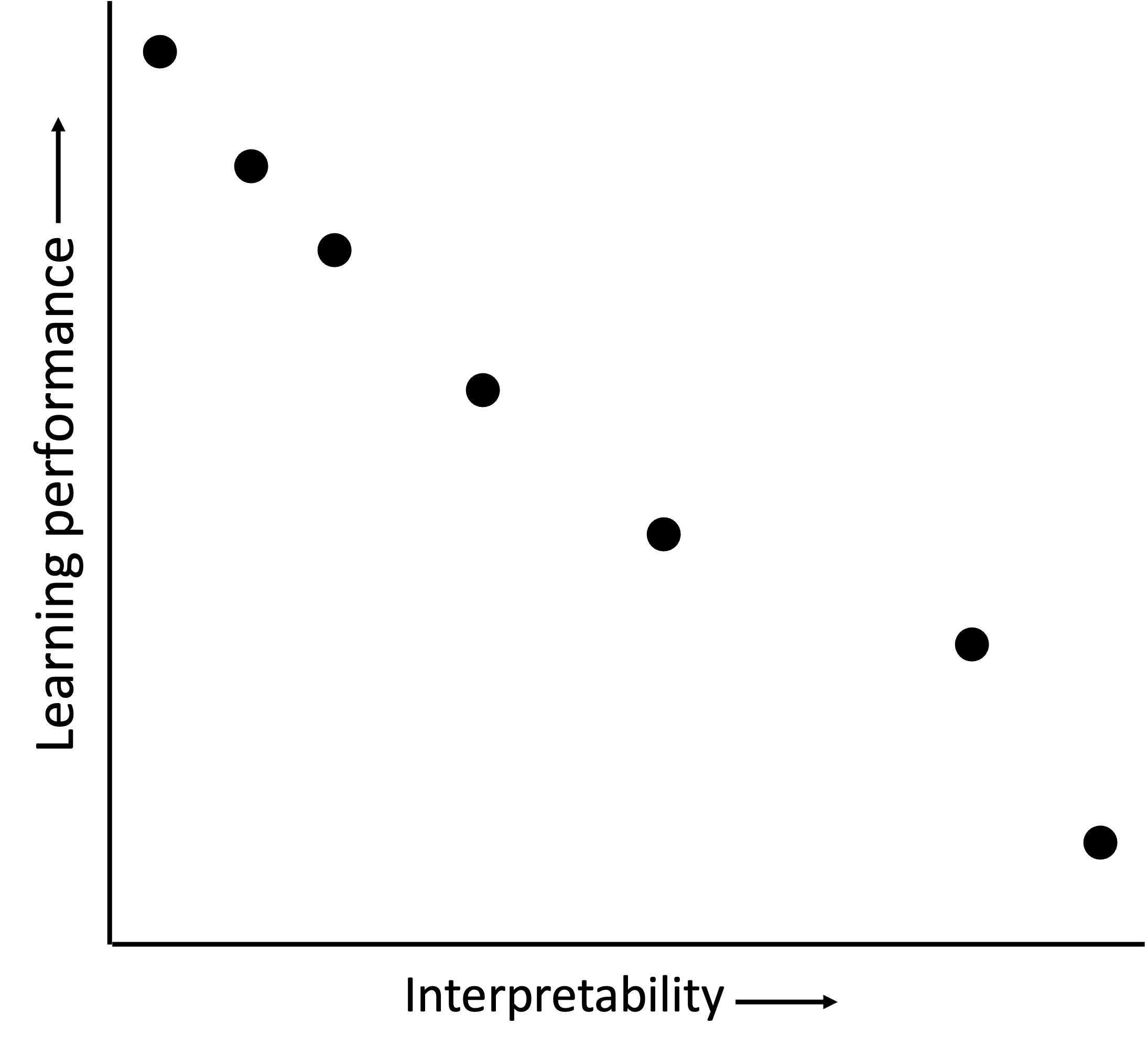}
         \caption{}
         \label{fig:accuracy-interpretability-pareto}
     \end{subfigure}
     \caption{Figure (a) represents a characterization of the two trade-offs in terms of a possible relation in the hypothesis space $\mathcal{H}$ between the set of accurate models $\mathcal{H}_A$ and the set of models that satisfy an additional formal constraint $\mathcal{H}_C$. The figure is inspired by Figure 1 in \citet{dziugaite2020enforcing}. (b) This figure, adapted from \citet{kearns2019ethical}, offers a toy illustration of the accuracy-fairness Pareto frontier, while Figure (c), adapted from \citet{gunning2019xai} (itself adapted from DARPA XAI presentation) offers a widely shared illustration of the accuracy-interpretability trade-off.}
     \label{fig:trade-offs}
\end{figure}

In terms of the set of accurate models, enforcing the constraint(s) may impose a predictive ``cost'' if $\mathcal{H}_C$ does not intersect with the set of accurate models: $\mathcal{H}_A \cap \mathcal{H}_C = \emptyset$ (see Figure~\ref{fig:accuracy-constraint}). These predictive costs can be quantified by comparing the performance of the best model in $\mathcal{H}_C$ with the models in the set of accurate models.

Of course, there can also be models that do satisfy these additional formal constraints, \textit{without} thereby suffering markedly in predictive accuracy (That is, $\mathcal{H}_A \cap \mathcal{H}_C$ is not necessarily empty). This is because, while models in the set of accurate models, $\mathcal{H}_A$, exhibit similar predictive performance, they can differ substantially in their formal characteristics that are relevant to fairness~\citep{coston2021characterizing,de2022algorithmic} and interpretability~\citep{semenova2019study,d2020underspecification}. Therefore, it would be a mistake to assume \textit{a priori} that enforcing formal fairness- or interpretability-motivated constraints will necessarily incur substantial or any predictive cost~\citep[See also][]{rodolfa2021empirical,rudin2019stop}. A promising area of research focuses on building tools and transparency documentations to help regulators and auditors better assess the normative significance of models within the set of accurate models~\citep[See][]{black2022multiplicity,coston2021characterizing}. With that in mind, let us examine in more detail how demands for fairness and interpretability have been formulated as enforcing constraints on learning. 

\subsection{Fairness as a constraint}

Concerns about the accuracy-fairness trade-off typically arise when we focus on the properties of predictive models~\citep{kearns2019ethical,corbett2017algorithmic}. Even within this relatively narrow scope, there is significant debate regarding how fairness should be formalized in terms of model properties~\citep{chouldechova2017fair,barocas-hardt-narayanan,corbett2017algorithmic}. However, we can largely set aside these disagreements because, in practice, many fairness measures can be framed as non-trivial parity constraints on the joint distribution $p(X,Y,\hat{Y},A)$, where $A$ represents membership in a protected group and $\hat{Y}=h(X)$ denotes the model's predictions~\citep{fazelpour2020algorithmic,barocas-hardt-narayanan}.

Achieving ``fairness'', from this perspective, can involve enforcing the relevant parity constraint(s) on the space of models~\citep{corbett2017algorithmic}. This might be done, for example, by adding fairness considerations as constraints in optimization~\citep[for example,][]{zafar2017fairness}, as penalty terms in the objective function~\citep[for example,][]{kamishima2011fairness}, or as group-specific thresholds for adjusting model behavior post-training~\citep[for example,][]{hardt2016equality}.\footnote{In the case of the latter post-processing approaches, the parity constraints that may restrict access to the set of accurate models are imposed not during optimization, but in the portion of learning that comes afterwards. This may be done, for example, by learning different decision thresholds that take group membership into account when translating conditional probabilities outputted by an unconstrained model (for example, risk of defaulting, if granted a loan) into binary classifications (for example, will default, if granted a loan)~\cite{liu2018delayed}. In doing so, these approaches can alter the \textit{predictive behavior} of the model in ways that no longer fits its unconstrained behavior. Insofar as learning these thresholds is also part of the broader learning of the algorithmic prediction model, these approaches fit the framework described here, even though the learning comes after machine learning optimization.} Broadly speaking, many of these algorithmic interventions restrict the set of available predictive models to those that satisfy the relevant parity constraints, which, as discussed above, can come at a cost to predictive accuracy.

\subsection{Interpretability as a constraint}\label{sec:accuracy-interpretability}
Although widely discussed in research and policy literature, the trade-off between accuracy and interpretability (or related terms like transparency and explainability) is less straightforward both formally and in terms of the potential value tensions it supposedly signifies. The trade-off is less straightforward in terms of value tensions because ``interpretability'' is typically not sought as an end in itself. Rather, its value is often tied instrumentally to serving various goals, such as increased understanding, enhancing user trust and decision support quality, improving deployment performance and safety,  fostering autonomy and procedural fairness, and ensuring public accountability~\citep{lipton2018mythos,murdoch2019definitions,krishnan2020against,jobin2019global,creel2020transparency,vredenburgh2021right,fleisher2022understanding}. 

Even if we assume that interpretability has instrumental value for achieving these goals, there are substantial disagreements about how to formally specify it and what specific properties make a model ``interpretable''~\citep{lipton2018mythos,krishnan2020against}. Once more, though, for the purposes of understanding the accuracy-interpretability trade-off, we can sidestep these more specific disagreements by broadly characterizing ``interpretability'' as a non-trivial constraint, whose enforcement restricts the set of admissible predictive models~\citep{dziugaite2020enforcing}. This characterization captures work in interpretable machine learning that formalize interpretability as some intrinsic model property (for example, low complexity or variable decomposability), thus \textit{directly} restricting the space of admissible models. It can also capture cases where demands for \textit{certain types} of\footnote{Our aim in this section is to understand those cases where concerns about accuracy-interpretability \textit{trade-off arise}. The claim is that in such cases, the demand for interpretability can be usefully characterized in terms of some constraint on the space of admissible models. This is distinct from the mistaken claim that \textit{all} interpretability or explainability methods constrain the model space. Many post-hoc explainability techniques are model-agnostic and can be applied without restricting the choice of the target model~\citep{murdoch2019definitions}. In those cases, concerns about accuracy-interpretability trade-off do \textit{not} arise (though, concerns remain about the reliability of these explanations~\citep[for example,][]{adebayo2018sanity,slack2020fooling,murdoch2019definitions}.} post-hoc explanations of model behavior \textit{indirectly} constrains the space of admissible models, because the additional insights required are more readily afforded by some models compared to others.\footnote{Perhaps given current state of knowledge, tools, or norms of practice about what counts as affording reliable understanding.}
For example, reliable quantification and communication of uncertainty (for example, confidence intervals) about model predictions can be critical for calibrated trust, information pooling, and safety in fields such as healthcare~\citep{bhatt2021uncertainty,kompa2021second}. Currently, techniques for uncertainty quantification are on a firmer theoretical basis for some models (for example, linear and logistic models) than others (for example, deep learning models and large language models)~\citep{murdoch2019definitions}.\footnote{Partly because the former group makes substantive assumptions about the underlying data-generating process---assumptions that can be justified by prior knowledge of the target phenomena, or may be erroneous simplifications. Of course, precisely because of this gap in theoretical understanding, uncertainty quantification and confidence calibration for deep learning and large language models is an active area of research, and so the situation might change~\citep{bhatt2021uncertainty,lin2023generating,detommaso2024multicalibration}.} In practice, then, those facing such demands may more readily use models that better afford such computations. 

Given this broad characterization, the accuracy-interpretability trade-off arises when no interpretable model exists within the set of accurate models. In such cases, one could quantify the ``price'' of interpretability in terms of the incurred loss of predictive accuracy~\citep{bertsimas2019price}. To capture this framing, in what follows we adopt a general understanding of the accuracy-interpretability trade-off that nonetheless tends to be operative in technical, philosophical and policy discussions: cases where we might need to sacrifice predictive accuracy to make gains in understanding of model behavior (broadly construed), in order to promote one of the aims for which interpretability, explainability, or transparency is taken to be a means~\citep[for example,][]{miller2019explanation,fleisher2022understanding}.

\subsection{Direct correspondence and a convenient division of labor}
How should we understand the relationship between the formal trade-offs in AI models and the varied impacts of deploying those models on corresponding values? If formal constraints, such as those related to fairness or interpretability, limit access to the set of accurate models and thereby reduce predictive accuracy, does this imply a corresponding sacrifice of decisional accuracy for the sake of supporting other substantive values? Conversely, if a decision-maker prioritizes decisional accuracy, does that mean they should prioritize proximity to the set of accurate models over imposing additional formal constraints?

A common interpretation of the trade-offs holds that there is a direct correspondence between model-level trade-offs and value tensions. For instance, \citet{babic2021beware} suggest that while decisional accuracy may take precedence in healthcare scenarios like diagnosis, interpretability---framed as a means to procedural fairness---should be prioritized in contexts like resource allocation. Taking this to have direct implications for model selection, they note that ``in [the latter] contexts, even if interpretable AI/ML is less accurate, we may prefer to trade off some accuracy, the price we pay for procedural fairness''~\citep[][p. 286]{babic2021beware}. In a similar vein,~\citet{london2019artificial} argues that in domains such as healthcare decisional accuracy and reliability matter more than understanding and explainability, and takes this to undermine proposed justifications for trading accuracy for interpretability at the model level.\footnote{London's position is more nuanced and we revisit it at the end of Sections~\ref{sec:validity} and~\ref{sec:expand-engagement}.} This direct correspondence interpretation is common in other discussions of the two trade-offs as well~\citep{london2019artificial,loi2021choosing,rahwan2018society}.

This prevalent interpretation of the trade-offs has important organizational, legal and policy implications. For example, it suggests a particular organization of responsible AI workforce, with a straightforward division of labor between technical and normative efforts. On the normative side, many philosophical and policy discussions assume that the formal trade-offs necessarily imply tensions between underlying values. Their role is then seen as resolving these value tensions, either by offering normative justifications for when, why, or to what extent some value should be prioritized in practice~\citep{london2019artificial,loi2021choosing,babic2019algorithms}, or by proposing participatory mechanisms that enable relevant stakeholders to make such determinations~\citep{rahwan2018society,lee2019webuildai}. Importantly, these justifications are taken to map directly on model choice. 

Conversely, on the technical side, researchers see their role as utilizing formal tools to guide decision-makers in navigating the trade-offs in model choice. For example, \citet{kearns2019ethical} see the role of formal tools as providing guidance for normative deliberation by mapping the accuracy-``fairness'' Pareto frontier---consisting of the set of models that cannot be improved in terms of one of these formal measures without incurring a loss on the other (see Figure~\ref{fig:accuracy-fairness-pareto}). 
Similarly, others offer guidance about how decision-makers, once they have decided that some sacrifice in predictive accuracy is justified, might go about choosing an appropriate ``interpretable'' model~\citep[for example,][]{molnar2022} (see Figure~\ref{fig:accuracy-interpretability-pareto}). And, these considerations about model choice are taken to map directly onto value prioritization. 

As will be discussed below, this interpretation has significant implications for many legal and policy debates in AI governance that center around value trade-offs. But is this interpretation of the value implications of trade-offs appropriate? In the next section, I turn to various considerations that complicate this view, and later, in Section~\ref{sec:discussion}, I return to the broader implications this shift in perspective holds for the division of labor, as well as for legal and policy debates.

\section{A Sociotechnical Perspective On Interpreting the Trade-offs}\label{sec:against}
Useful as it might be for motivating the discussion of the formal trade-offs, the characterization of AI-based decision-making in the previous section involves significant, and potentially problematic simplifications. 
Specifically, researchers have argued against framing the value implications of AI models solely in terms of model properties, in isolation from their context of development and use~\citep{selbst2019fairness,fazelpour2020algorithmic,herington2020measuring,raji2022fallacy,suresh2019framework}. Instead, there have been calls for adopting a sociotechnical perspective that encompasses not only the properties of AI models, but also other technical, psychological, organizational, and social factors that shape the life-cycle of AI-based decision systems, and ultimately their societal impacts. 

In this section, I approach the trade-offs from this sociotechnical perspective, discussing three sets of considerations necessary for bridging the gap between formal, model-level trade-offs and their practical value impacts. As we will see, addressing these sociotechnical considerations shows that we \textit{can} achieve better outcomes than the formal trade-offs might suggest. Beyond diagnosing why the common interpretation of these trade-offs is mistaken, the sociotechnical approach also provides practical guidance on how to achieve these improved epistemic and ethical possibilities. This, in turn, offers compelling reasons for evaluating and governing AI from a sociotechnical perspective.\footnote{Thanks to an anonymous reviewer for the suggestion.} 

\subsection{Validity and relevance}\label{sec:validity}
To take the two formal trade-offs as directly having the purported normative implications, we must assume that the formal constructs involved aptly track the relevant values. But this assumption can fail, when invalid operationalization, measurement, and estimation procedures and practices result in disconnects between formal constructs and the epistemic, ethical, or legal aims and values that those constructs are supposed to capture. 

Consider first the notion of ``predictive accuracy'', which is implicated in both trade-offs. In many prediction-based decision settings, institutions care about inferring outcomes that are ambiguous, latent, and contested, such as being a patient with ``severe healthcare needs''~\citep{obermeyer2019dissecting}, a child ``at risk''~\citep{saxena2020human}, or a tweet containing ``hate speech''~\citep{waseem2016you}. Let us refer to this outcome that we practically care about as $Y_c$. Rendering $Y_c$ suitable for machine learning application often requires selecting a simplified and unambiguously measurable proxy outcome, $Y$. This is not a trivial task, and 
often involves various value judgments. In practice, $Y$ might not be a valid proxy for $Y_c$ for a variety of reasons, ranging from lack of construct validity in operationalization to challenges of bias and validity in measurement, estimation, and aggregation~\citep{fazelpour2021algorithmicbias,de2018learning,kleinberg2018human,jacobs2021measurement}. 

Importantly, for our purposes, the accuracy-fairness trade-off can arise when the extent of the disconnect between $Y$ and $Y_c$ is not evenly distributed across the population of interest~\citep{de2022algorithmic}.\footnote{Intuitively, this can be thought of as potentially resulting in a statistical dependence between $Y$ and some protected attribute $A$ not because one exists in reality, but as a result of differential validity in our categorization and data collection practices.} That is, when $Y$ exhibits differential validity across groups, optimizing for predictive accuracy can undermine some fairness-motivated parity constraints (and vice versa). In a salient example of such differential validity in healthcare, the disconnect between a patient's ``healthcare need'' and its operationalization in terms of ``healthcare expenditure'' particularly harmed Black patients, for whom expenditure was particularly an unreliable proxy due to a variety of factors including justice-related ones~\citep{obermeyer2019dissecting}. Similar issues arise when our measurement, estimation, and aggregation techniques exhibit differential validity across protected groups~\citep{coston2023validity,jacobs2021measurement}.\footnote{For example, if our data collection and aggregation practices make it more likely that tweets from certain marginalized groups are more likely to be labeled as ``toxic speech''~\citep{davani2022dealing}.}

In such cases, it would be a serious error to take the accuracy-fairness trade-off as an inescapable fact and to focus our deliberative efforts simply on selecting among the set of Pareto optimal solutions. Rather, we need to first examine the design and development choices that lead to the underlying issue of differential validity, and if possible address this underlying issue by adopting better sociotechnical practices in problem formulation, operationalization, and data collection~\citep{hellman2020measuring,obermeyer2019dissecting,kleinberg2018discrimination,fazelpour2021diversityml}. Put another, instead of taking the formal trade-off as an expression of tension between corresponding values, we can use the formal trade-off as a tool for interrogating the aptness of those upstream design choices, and identifying and addressing potential epistemic-ethical problems that undermine not only fairness, but also accuracy of the sort decision-makers actually care about. This re-framing allows us to recognize that we \textit{can} have things better than a simple interpretation of the formal trade-offs might imply.

Similar relevance concerns arise for formal notions of ``interpretability''. Currently, there exists a considerable disconnect between the underlying values for which ``interpretability'' is sought (for example, trust, improved decision quality, safety, recourse, ...) and the technical operationalizations of the term (for example, in terms of model complexity, variable decomposability, model type, ...)~\citep{krishnan2020against,lipton2018mythos}. Indeed, in some cases, formal operationalizations appear to undermine the very values they are meant to support, such as when individuals lose their trust in algorithmic predictions upon seeing that they originated from a simple decision tree, whose lack of complexity was meant to invite user trust~\citep{lu2019good}.\footnote{Of course, there can be ways of addressing this loss of trust in the simpler, less accurate model (for example, via stakeholder participation, education, and deliberation). The point here is simply that, at least in some cases (for example, domains, audience, etc.), interventions that constrain space of models in the name of values that motivate interpretability (for example, trust) can come at a needless cost, and those underlying values can be supported more effectively, with more accurate models.} When a measure of ``interpretability'' lacks instrumental relevance in this way, it is unclear why we would want it, especially if attaining it comes at an avoidable cost to predictive accuracy. What is more, even if interpretability (in some sense) is \emph{a} means to those underlying values, it may not be the only, or even the most effective, means in context. Indeed, in some cases, ensuring \textit{accuracy} across diverse datasets may offer a better instrumental approach. In healthcare settings, for example, \citet{london2019artificial} proposes a number of alternatives, including this type of accuracy-based pathways, for achieving some of the aims for which interpretability is sought (for example, safety)~\citep[see also][]{krishnan2020against}.\footnote{We will return to London's view in Section~\ref{sec:expand-engagement}. For a critique of London's view see \citet{grote2023allure}.}

As will be argued below, this is, of course, not to say that interpretability---in the broader sense defined in Section~\ref{sec:accuracy-interpretability} and suitably specified---cannot be critical for supporting key epistemic, social, and ethical aims in many settings~\citep[see also][]{murdoch2019definitions,vredenburgh2021right,creel2020transparency,grote2023allure}. But the \emph{practical} implications of a potential accuracy-interpretability trade-off is dubious, if the operationalized measure of interpretability lacks instrumental relevance (and necessity) for promoting the values that animate concerns about interpretability. Closely inspecting the validity and instrumental relevance of the formal constructs can uncover normatively significant disconnects and alternatives. In many cases, addressing these underlying issues can enable us to robustly promote both sets of relevant values in ways that remain outside the purview of the narrow formal characterization of the trade-offs. 

\subsection{Compositionality}\label{sec:compositionality}
In many cases of social concern, algorithmic tools are not stand-alone decision-makers, but  function as part of broader decision-making systems. Examples include healthcare~\citep{kompa2021second}, child welfare services~\citep{de2020case}, loan approvals~\citep{paravisini2013incentive}, and fact-checking~\citep{guo2022survey}, where AI models assist (rather than replace) human experts. That is, in many highly publicized cases of ethical concern about the unfairness or opacity of predictive models, these models function as decision \textit{supports}. In such cases, the focus should shift from the AI model's isolated performance to how its integration can improve the overall decision quality, especially compared to the status quo of unaided human experts~\citep[See][]{green2019disparate}. It thus becomes crucial to examine how the properties of AI models (accuracy, fairness, interpretability, or their trade-offs) contribute to the broader decision-making systems in which the AI models are embedded.\footnote{In this section and the next, the discussion will primarily focus on decision-making systems that involve human experts, given the prevalence of such cases as well as their policy relevance, as a result of policy demands for human oversight, human-in-the-loop, and more. However, the broader issues of compositionality (this section) and backward compatibility (next section) pertain to many other socially consequential systems that are composed of multiple, interacting algorithms. Examples include content recommendation systems~\citep{twittergithub} and fact-checking~\citep{guo2022survey} systems on social media.} 

This shift of unit of analysis from individual components to composed groups has significant epistemic and ethical implications for evaluation and design. Works on \textit{complementarity} and \textit{collective intelligence} in teaming and group performance illustrate the divergence between evaluating and designing (or selecting) at the individual versus the group level~\citep{page2019diversity,steel2018multiple,riedl2021quantifying}. This body of theoretical and empirical research demonstrate that across various tasks---such as classification, prediction, problem-solving, and medical diagnosis---group performance is not strongly correlated with individual performance, and that suitably created groups (for example, with complementary skills, or appropriate information pooling) can not only outperform the best performing individuals, but may not even include those individuals~\citep{page2019diversity,laughlin2006groups,woolley2010evidence,aggarwal2019team,kurvers2016boosting,krause2010swarm,bahrami2010optimally,bang2017making}.\footnote{This realization also underpins various \textit{ensemble learning} methods in machine learning.} As these works show, especially in complex tasks, a group's performance hinges, not simply on individual performance, but on how members' cognitive abilities and limitations (information, background knowledge, decision heuristics, biases, ...) interlink. For instance, one member's different perspective can offer a breakthrough on a problem that hinders another. Conversely, this synergy is lost, if members' strengths and weaknesses are too similar. Crucial too are appropriate task allocation (who should do what cases) and effective information pooling, communication, and integration (whose opinions should weigh how much). 

Compared to the long history of research in these other fields, discussions of designing for group-level benefits are relatively nascent in the context of AI-assisted decisions. Nonetheless, these recent works show that substantial gains can be made at the level of the human-AI systems, if we incorporate complementarity and other group-level considerations as explicit design and evaluation criteria. Specifically, these works have explored issues such as: characterizing the type and extent of variability in individual performance required for complementarity~\citep{donahue2022human}; identifying the mechanisms that could give rise to such differences between human experts and AI models (for example, access to different information)~\citep{rastogi2022unifying}; 
designing optimization objectives for models that suitably complement human experts~\citep{mozannar2020consistent}; 
developing optimal task allocation schemes~\citep{madras2018predict}; 
designing models and tools that facilitate effective information pooling~\citep{van2020interpretable,bansal2019beyond}; 
and designing deferral tools to ask for second opinion (for example, on cases where AI systems are uncertain, or when experts are known to be better)~\citep{kompa2021second,raghu2019direct}.\footnote{Some of these works can also be seen as computational implementations of group-level norms such as \textit{contestability} considerations~\citep[for example,][]{anderson2006epistemology,landemore2020open}.} 

These works underscore that properly evaluating AI model properties and trade-offs depends on understanding their contribution to the qualities of the broader decision-making system. Many of the above techniques can result in reduced model accuracy, but improve accuracy at the system-level. Notably, in order to achieve these, in some cases, model accuracy is traded for improved understanding. 
This might be done to help experts construct suitable mental models of the AI behavior~\citep{bansal2019beyond} or be better placed to combine AI outputs with their own knowledge~\citep{caruana2015intelligible}. More generally, interpretability, as defined in Sec~\ref{sec:accuracy-interpretability}, can also alleviate experts' distrust of AI models, promoting better informational uptake and integration~\citep{zerilli2022transparency,grote2023allure}.\footnote{Interestingly, some post-hoc explanations (for example, local saliency explanations) have been shown to have \textit{detrimental} impact on the quality of AI-informed decisions by increasing experts reliance on (and reducing their vigilance about) AI output~\citep{bansal2021does}. But, caution is needed when interpreting such findings. 
We must distinguish between the values of interpretability and the effectiveness of particular interpretability-seeking techniques in achieving those values; 
the fact that some techniques can result in misleading explanations does not undermine the value of explanations~\citep{lombrozo2011instrumental} or improved understanding~\citep{grimm2012value} in general. Moreover, even when sometimes unjustified, increased reliance overall on AI can still be a net epistemic good, depending on the status quo of unaided human decision making.} 
Thus, a perceived compromise at the model-level can in fact translate into an advantage at the broader system-level.

The study of ``fairness under composition'' reveals similar insights~\citep{dwork2018fairness,wang2021practical}. As~\citet{chouldechova2020snapshot} note
``often the composition of multiple [formally] fair components will not satisfy any fairness constraint ... [while] the individual components of a fair system may appear to be unfair in isolation''~(p. 86).  As works in social services and criminal justice settings further show, the integration of AI predictions in human decision-making can lead to complex dynamics, where the distribution and type of inaccuracies and biases from AI models and human decision-makers may compound, offset, or beneficially complement each other~\citep{de2020case,green2019disparate,skeem2020impact,donahue2024two,mclaughlin2022fairness}. Here too, moving beyond a model-centric view is essential to recognize potential (individual vs. system) divergences and understand how principled model-level ``losses'' along a dimension can result in system-level gains along that same dimension.

When predictive models compose just one \textit{part} of broader decision-making systems, then, it is a \emph{category mistake} to focus on their properties apart from those of the other parts (for example, user characteristics and capabilities, organizational norms). In such cases, the simple interpretation of the two trade-offs is not only misguided, it can also distract us from other promising epistemic and ethical considerations. In Section~\ref{sec:discussion}, we will discuss the philosophical, legal, and policy implications of this change of perspective. 

\subsection{Deployment dynamics}\label{sec:dynamics}
To responsibly integrate AI models into social decision pipelines, it's essential to understand more than just model properties within a static testing environment. It also requires a grasp of the mechanisms governing the dynamic interactions of AI outputs with their organizational and social embeddings. Understanding these dynamics is key for properly contextualizing model-level trade-offs. What appears as a sacrifice of one desideratum for another from a static perspective could be necessary for ensuring sustained gains in the former in the longer term. 

Consider first how our thinking about fairness and accuracy can be informed by better understanding social dynamics.\footnote{For a similar discussion on accuracy-fairness see \citet{de2022algorithmic} and \citet{fazelpour2021algorithmic}} Findings from workforce diversity research offer a salient example. These works show that the benefits of increased team diversity---often understood in this literature to be maximal at demographic parity---is not linear, and depends, among other things, on how diverse a team \emph{initially} is~\citep{phillips2017real,bear2011role,post2015women}. That is, while moving towards demographic parity can have negative effects on team performance in highly homogeneous teams, 
it can improve performance in groups that are (to some extent) diverse~\citep{steel2018multiple}. 
This happens because the changed team composition alters the dynamics of the environment in which team members collaborate, in ways that allow the team to better elicit and integrate different perspectives and capabilities. 
Accordingly, short-term utility losses for the sake of parity on fairness-related grounds can be offset by robust, long-term utility gains (understood as improved team performance on a variety of tasks). 
In other cases, making short-term gains in fairness at the cost of accuracy can result in compounding injustices in the longer term~\citep{fazelpour2021algorithmic}. 
Take, for example, fairness-related disparities in lending that arise because of conditions affecting the repayment ability of members of certain disadvantaged groups. 
In such cases, achieving fairness-related parities requires reducing predictive accuracy, resulting in granting loans to individuals who may not be able to repay them. 
Yet, the resulting disproportionate defaults can widen existing discrepancies (for example, due to its impacts on other opportunities via damaging credit scores)~\citep{liu2018delayed,d2020fairness}.\footnote{As \citet{d2020fairness} note, whether the net effect of our action---that is, granting more loans to members of disadvantaged groups vs. widening the credit gap between the groups---is positive depends on our values. But, as they also note, we will not be in a position to \emph{evaluate} this tension, if we simply focus on fairness as a static property of a model.} 
That is, making short-term gains in fairness at the cost of accuracy can potentially result in compounding injustices in the longer term.

Examining these dynamics is also crucial for improved reasoning about longer term predictive accuracy and its relationship with interpretability. 
Issues around updating AI models offer an apt example here. 
Updating can improve predictive accuracy by leveraging increased data, especially from post-deployment observations; it is also crucial for addressing dynamic distribution shifts (for example, due to environmental changes or strategic behavior of decision subjects) that can undermine predictive capabilities~\citep{babic2019algorithms,zrnic2021leads}.

But the accuracy of the updated AI model with respect to a static dataset is not the only consideration. 
We also need to consider the dynamic interactions between the updated model and the rest of the decision-making system~\citep{bansal2019updates,srivastava2020empirical,bansal2019beyond}. 
For example, while enhancing a model's overall accuracy, updates can degrade accuracy in cases where users had come to justifiably rely on the model~\citep{bansal2019updates}. These unexpected shifts can hurt the overall human-AI performance, and undermine trust, in ways that negatively affect experts' long-term adoption of AI tools~\citep{dietvorst2015algorithm}. Updating that ensures sustained improvements can thus require potential losses to model accuracy for the sake of backward compatibility~\citep{bansal2019beyond,bansal2019updates,srivastava2020empirical}. For example, some interventions for ensuring the backward compatibility of the updated models constrain the space of learning to those models that are (to an extent) consistent with the understanding (or mental model) that the user had formed of the algorithm. In this way, these interventions can make the models more interpretable (in the sense discussed in Section~\ref{sec:prediction-based}), resulting in performance gains at the system level over time. 

More generally, AI models are often embedded in complex epistemic practices~\citep{creel2020transparency,fleisher2022understanding}. Appreciating the dynamic and evolving nature of these practices should inform our understanding of the relation between accuracy and interpretability. As noted by \citet{rudin2019stop}, for example, interpretable models that exhibit lower accuracy on a static dataset (compared to a ``blackbox'' model) may afford better opportunities for understanding and iteratively refining the knowledge discovery process, thus improving accuracy in the longer run~\citep[See also][]{murdoch2019definitions}. 

Overall, then, static properties of predictive models do not tell the full story about what we stand to gain or lose by adopting the models in complex, changing environments. 
What might appear to be an inescapable trade-off from a static perspective may not be one, once we adopt a broader and longer term perspective. 
Similarly, justifications developed in the abstract about how such trade-offs should be navigated may provide hardly any epistemic or ethical reassurance in complex social systems.  

\section{Broader Implications}\label{sec:discussion}
Having discussed each of the considerations separately, let us bring out some of the shared lessons.

\subsection{Expanding the scope of normative engagement}\label{sec:expand-engagement}
The previous section demonstrated how the direct correspondence interpretation of formal trade-offs can lead to misplaced justifications and missed opportunities for achieving better epistemic and ethical outcomes. To achieve these improved possibilities, it is crucial to expand the scope of epistemic and ethical normative engagements to include the full range of considerations that influence the validity of model properties (Section~\ref{sec:validity}) as well as the cognitive, organizational, and social factors that shape the situated impacts of AI models in real-world settings (Sections~\ref{sec:compositionality} and~\ref{sec:dynamics}). 

Importantly, these factors must be addressed in an integrated and coherent manner. For example, in his insightful discussion of the accuracy-interpretability trade-off, \citet{london2019artificial} argues that, in medical practice, \textit{decisional accuracy} and reliability take precedence over interpretability. Moreover, as mentioned in~\ref{sec:validity}, London also questions the instrumental value of interpretability, noting that many of the goals associated with interpretability could be achieved through alternative interventions that do not compromise accuracy. Should we, therefore, abandon the pursuit of interpretable models when they reduce \textit{model accuracy}? Not necessarily. This because London's discussion overlooks key considerations relevant to medical settings, where AI tools are used as decision supports to human experts in dynamic environments. As discussed in Sections~\ref{sec:compositionality} and~\ref{sec:dynamics}, in many such cases, ``sacrificing'' model-level accuracy for interpretability can result in system-level gains in decisional accuracy that are also sustainable in changing environments. 

To identify and achieve these more meaningful gains, our normative engagement with trade-offs must encompass the full set of considerations in an integrated way.\footnote{Of course, not all considerations may always be directly relevant in every case. For instance, in some cases, it may be justifiable to infer decision properties directly from model properties, and abstracting away issues like compositionality. However, deciding whether such abstraction is appropriate is itself a contextual and value-laden judgment, best approached from a sociotechnical perspective~\citep{selbst2019fairness}.} Relying on normative reasoning that does not engage with these broader factors may not only result in missed opportunities but may also impose unnecessary costs on relevant stakeholders. This does not imply that value trade-offs—such as those between accuracy, fairness, and interpretability—will never arise. On the contrary, this perspective brings into focus other, often overlooked, value tensions. 

For example, enhanced participation and contestation can improve group decision-making in various stages of the AI lifecycle, from operationalization and measurement~\citep{martin2020participatory,lee2019webuildai} to human-AI teaming~\citep{kompa2021second,raghu2019direct}. Yet, such processes can also result in delays or undermine group cohesion and decision quality~\citep{fazelpour2021diversityml,dobbe2020hard}. Similarly, when addressing issues of compositionality, we encounter the well-known diversity-stability trade-off from complex systems theory, where too much variability—a prerequisite for complementarity—can destabilize the system and degrade performance~\citep{page2019diversity,eliassi2020science}.

The crucial point, then, is that better design and governance possibilities as well as these practically relevant debates risk being overlooked if normative deliberation remains focused solely on formal trade-offs, which may not have the assumed value significance.

\subsection{``Business necessity'', transparency, and implications for law and policy}
This sociotechnical perspective on interpreting the value implications of formal trade-offs has significant legal and policy implications. Many debates in these domains hinge on careful characterizations of trade-offs between competing social objectives. The adjudication of disparate impact cases in the U.S. legal system offers a particularly relevant example. In these cases, the central question is not simply whether a practice (for example, a hiring policy) has disproportionate adverse impacts on a protected group, but whether there is a legitimate, competing objective (for example, business aims) that cannot be adequately promoted by alternative means, and so justifies the practice despite its impacts~\citep{sep-discrimination}. As Barocas and Selbst note, evaluating such ``business necessity'' justifications and the purported trade-off between business utility and fairness is ``the crux of disparate impact analysis''~\citep[][p. 702]{barocas2016big}. 

In the context of AI-based decision-making, the tension between social values like business utility and fairness has increasingly been framed in terms of the formal accuracy-fairness trade-off~\citep{kleinberg2018discrimination,mayson2018bias,barocas2016big,ho2020affirmative,valdivia2021fair,kirat2023fairness}.\footnote{Model accuracy is often associated with business utility because it is considered predictive of task-relevant outcomes~\cite[See for example, ][]{barocas2016big}.} 
Indeed, some scholars view this framing of value tensions in terms of formal trade-offs as a positive development. For instance, Kleinberg et al. argue that algorithms provide a means to ``precisely quantify trade-offs among society's different goals,'' enabling more rigorous evaluations of whether practices can be justified on business necessity or similar grounds~\citep[][p. 151]{kleinberg2018discrimination}. As a result, they suggest that, in legal contexts where competing social objectives are at play, algorithms should be viewed not merely as potential threats but as tools for advancing equity by bringing transparency to value trade-offs.\footnote{\citet{kleinberg2018discrimination} also include debates about affirmative action in their discussion, where diversity serves as a social objective, potentially competing with accuracy. As they suggest, ``with the algorithm it is now possible to see whether there is a tradeoff between diversity and academic performance ..., and if so, trace out what that tradeoff curve looks like''~\citep[][p. 151]{kleinberg2018discrimination}.} 

Of course, as discussed in Section~\ref{sec:validity}, serious ethical concerns can arises if decision-makers simply accept the formal trade-off at face value, and neglect how specific design choices shape the trade-off itself. Kleinberg et al. adopt a more nuanced stance, emphasizing that the transparency of the formal trade-off is valuable insofar as there exist auditing mechanisms that allow for the scrutiny and experimentation of the design choices that shape the trade-off. They identify three key areas for examination: (i) the operationalization of the target outcome; (ii) the inclusion of features in the dataset; and (iii) the objective optimized during model training~\citep[][p. 152]{kleinberg2018discrimination}. Recent efforts to standardize transparency documentation, such as data sheets and model cards, are important steps toward making these design choices visible and open to evaluation~\citep{winecoff2024}.

A sociotechnical perspective, as developed here, substantively re-frames even this more nuanced approach to evaluating business necessity (or other value trade-offs) in AI-based decision-making. The discussions in Sections~\ref{sec:compositionality} and~\ref{sec:dynamics} make it clear that we must be skeptical of Kleinberg et al.'s claim that algorithms allow us to  ``precisely quantify trade-offs among society's different goals''. To properly connect formal trade-offs to broader social goals, it is crucial to understand how model properties interact with other components of the decision-making system (for example, human users) in dynamic, real-world environments. This re-framing not only alters the types of justifications required but also calls for new forms of transparency documentation that go beyond existing tools, which primarily focus on detailing dataset and model properties.\footnote{For a comprehensive list of existing transparency documentation, see \citep{winecoff2024}.}

Consider the argument in Section~\ref{sec:compositionality} that the integration of AI in organizational decision pipelines impacts decision quality (for example, accuracy or fairness) in ways that cannot be reliably anticipated by examining formal model properties alone. Rather, to evaluate the quality of AI-informed decisions, and to justify the use of AI, it is essential to understand how the capabilities and limitations of AI models relate to the expertise of human professionals---whether they overlap, reinforce, offset, or beneficially complement each other---and the psychological and organizational factors that shape how experts integrate AI predictions into their decision-making processes.\footnote{Ideally, these anticipatory evaluations should be further supported by empirical studies measuring how the integration of AI models impacts the overall quality of AI-informed decisions across meaningful organizational outcomes.} Yet, despite the critical role of human experts in institutional uses of AI (for example, criminal justice, social welfare, and healthcare), and despite policy documents that mandate human involvement and oversight, current transparency and risk management tools fail to make this crucial information about human-AI collaboration legible to evaluators and auditors.\footnote{For a list of policy documents requiring human involvement and oversight, see~\citep{green2022flaws}.} Future normative work is required for the appropriate development of such transparency documentation and their underpinning. These efforts can take inspiration from existing frameworks used to capture human factors in managing the risk of medical devices~\citep{pelayo2021human,food2016applying}.

Indeed, focusing on the design of better human-AI systems---rather than focusing solely on individual models---could change how we evaluate and justify the use of AI even more radically. 
Currently, standard approaches to AI evaluation and design tend to focus on tasks where human experts already perform well, rather than on tasks where experts need the most assistance~\citep{blagec2023benchmark}. Even in those tasks, AI models are typically designed and evaluated based on their ability to \textit{compete} with human experts, rather than complement or collaborate with them~\citep{haenssle2018man,pham2021ai}.\footnote{This approach is also mirrored in journalistic headlines, such as ``AI beats radiologists'' or ``AI is better than dermatologists,'' reflecting the broader competitive and individualistic framing.} In these cases, not only is it a distortion to focus on model-level ``gains'' or ``losses'' (as opposed to their system-level upshots) in framing deliberations about value trade-offs; it can also be misleading to engage in such deliberations without a clear empirical grasp of the psychological and organizational realities that shape where AI assistance can most benefit experts and which populations benefit most from that assistance. The focus on formal trade-offs as direct expressions of social value tensions not only obscures the critical influence of individual actors (for example, experts) on AI-informed outcomes but also conceals the consequences of key organizational decisions about where and how to integrate AI (for example, whether it actually supports meaningful expert involvement and oversight). The concept of ``agency laundering'', as introduced by \citet{rubel2019agency}, can offer a powerful lens for future normative work interrogating the concealed responsibility and accountability in these individual and organizational choices. 

Similarly, as emphasized in Section~\ref{sec:dynamics}, focusing on static model trade-offs provides an incomplete and potentially misleading understanding of the value consequences of adopting AI models in dynamic and evolving environments. For example, a practice that was once justified under business necessity may not continue to be so as conditions change or as models are updated. So, the task of justification in these cases cannot be limited to evaluating static model trade-offs. It requires documenting and normatively defending the strategies used to manage such dynamic effects. The existing lack of documentation on how these changes are managed, particularly in light of other components of the organizational context, creates another layer of opacity that must be addressed by future efforts.\footnote{Recent work on ``reward reports'' for reinforcement learning algorithms by \citet{gilbert2023reward} offers a useful inspiration for future development in this area. \citet{babic2019algorithms} provide a thoughtful discussion of policy and regulatory challenges around updates.} 

As mentioned above, in re-framing the scope of normative engagement, the sociotechnical perspective also brings into view \textit{other value tensions} that should inform the evaluation and justification of alternative design choices. Consider, for example, Kleinberg et al.'s framing of the AI-afforded transparency and precision as an unmitigated good. However, this view overlooks the positive value (and sometimes necessity) of \textit{constructive ambiguity}---a tool organizations deliberately use to manage diverse stakeholder interests~\citep{eisenberg1984ambiguity}. In allocation contexts, ambiguity in defining outcomes of interest (for example, ``good applicant'') can facilitate conflict resolution among different stakeholders by accommodating their varied interpretations, often by allowing room for discretion at the decision-making stage~\citep{elster1992local}. In this way, constructive ambiguity can play a vital role in supporting procedural fairness and fostering stakeholder trust. Viewed in this way, evaluating different operationalizations is not just a mater of inspecting effects on formal trade-offs, but also attending to potentially different ways that construct precision can interact with discretion, inter-stakeholder cohesion, and procedural fairness. More generally, from a sociotechnical perspective, evaluating design choices requires assessing not only their effects on formal model properties (for example, accuracy, fairness, trade-offs) but also their broader organizational and social impacts.

By offering a more comprehensive view of how algorithms function within their organizational and social contexts, the sociotechnical approach enables a deeper evaluation of whether a given policy is genuinely necessary for advancing specific objectives or whether alternative strategies could achieve similar or better outcomes without sacrificing other values and objectives. Naturally, identifying, mapping, and carefully analyzing these key decisions and their value implications demands a broad range of expertise and a new model of interdisciplinary collaboration for responsible AI evaluation and governance. It is to this that I now turn.

\subsection{The meaning of ``AI talent'' and the need for interdisciplinarity}
As mentioned in Section~\ref{sec:prediction-based}, the direct correspondence interpretation is often accompanied by an implicit division of labor between technical and normative efforts. In particular, it suggests that once researchers, stakeholders, or policy-makers decide which of their values should be prioritized in a given context, the rest is technical work. Notably, these judgments about the relevance and priority of values are often made without close engagement with the various choices involved throughout the design, development, and deployment processes of AI. For example, arguments about the priority of accuracy over interpretability are offered by considering the demands of different medical applications, as opposed to how AI models are designed for and embedded into those applications. 

This conception of the disciplinary division of labor may also underpin recent governmental orders and directives that aim to promote responsible AI. For example, in its recent efforts to promote responsible AI, the United States government has launched an \textit{AI Talent Surge}.\footnote{See, for example, the directives by the Biden-Harris Administration.} Importantly, however, the notion of the ``AI talent'' that is meant to help ``federal agencies to responsibly leverage AI'' is understood exclusively in technical terms in terms of data science and tech talent.\footnote{See \url{https://ai.gov/apply/}} But, if the considerations above are on the right track, this is serious misconception. 

As discussed above, the prevalent interpretation of the formal trade-offs can mislead our normative deliberation and result in costly disconnects between implementations and justifications that are meant to ground them. A sociotechnical perspective not only offers a safeguard against misinterpreting the trade-offs (and model properties more generally); it can also offer improved epistemic and ethical opportunities. But taking advantage of these opportunities requires serious interdisciplinary collaborations. Advancing our understanding of any of the considerations above (for example, operationalizing social phenomena, human-AI complementarity and collaboration) requires drawing on significant domain expertise as well as the knowledge and methodologies from a variety of disciplines, including those in the humanities and social sciences~\citep[For related points see][]{fazelpour2021diversityml,stinson2024feeling,rudin2019stop}. 

Of course, in taking an interdisciplinary approach to responsible AI seriously, we need to address many challenging questions. For example, at an interpersonal level, diverse and interdisciplinary teams often face communicative problems~\citep{o2013enhancing,page2019diversity}. What type of upskilling is needed to address these challenges in the context of responsible AI design and governance? 
At an organizational level, successful interdisciplinary collaboration appears to require factors such as sustained leadership support, egalitarian power relations between groups, and more~\citep{phillips2017real}. What interventions and incentives can effectively realize these factors in the responsible AI ecosystem? At a structural level, what are the appropriate organizational structures for effective interdisciplinarity (for example, should we have teams of mostly interdisciplinary individuals or disciplinary clusters with interdisciplinary bridges; what type of publication and conferences venues are most suitable)?~\citep{leydesdorff2008betweenness} The first step towards making advances on these questions is to appreciate that responsible AI governance cannot simply be achieved by the vision of disciplinary division of labor underpinning the prevalent interpretation. If ``AI talent'' is key to responsible innovation and design, then that talent needs to be understood in an interdisciplinary way. 

\section{Conclusion}\label{sec:conclusion}
With the increasing social use of AI, we need to carefully examine the epistemic and ethical questions posed by these technologies. Taken together, the three sets of considerations discussed above can provide a lens that can assist those involved in evaluating, designing, and investigating these systems. While the focus here is on the two formal trade-offs in AI-based decision-making, the considerations may be applicable to formal trade-offs in other normatively significant and policy-relevant scenarios. This is because turning a policy scenario into one amenable to formal treatment often requires abstractions and assumptions about the broader sociotechnical context that often tend to be neglected in interpreting. The discussion can thus potentially offer something to researchers interested in those other scenarios. Hopefully, the considerations above allow us to leverage these formal trade-offs in our practical deliberations, without allowing a misleading interpretation of them punish our normative aspirations in a negative sense.

\section*{Acknowledgments}
I'd like to thank audiences at the ANU Machine Intelligence and Normative Theory lab, the Statistical Society of Canada's 2023 Annual Meeting, National Institute of Standard and Technology's AI Metrology Colloquia Series, Machine Wisdom Workshop, and Harvard's Edmond \& Lily Safra Center for Ethics for valuable comments and discussion. Many thanks also to the three anonymous referees for their constructive feedback and critique. This research was supported by Schmidt Sciences AI2050 Early Career Fellowship.

\bibliographystyle{apalike}
\bibliography{refs.bib}
\end{document}